# The SU(3) running coupling from lattice gluons

*UKQCD Collaboration* - presented by C. Parrinello[*]

Department of Physics and Astronomy, The University of Edinburgh, Edinburgh EH9 3JZ, UK.

We provide numerical results for the running coupling in $SU(3)$ Yang-Mills theory as determined from an analysis of lattice two and three-point gluon correlation functions. The coupling is evaluated directly, from first principles, by defining suitable renormalisation constants from the lattice triple gluon vertex and gluon propagator. For momenta larger than 2 GeV, the coupling is found to run according to the 2-loop asymptotic formula. The influence of lattice artifacts on the results appears negligible within the precision of our measurements, although further work on this point is in progress.

## 1. INTRODUCTION

The running coupling $\alpha_s(q)$, where $q$ is a momentum scale, is a fundamental QCD quantity providing the link between the low and high-energy properties of the theory. Given a renormalisation scheme, $\alpha_s(q)$ can be measured experimentally for a wide range of momenta. A precise determination of $\alpha_s(q)$ (or equivalently of the scale $\Lambda$ determining the rate at which $\alpha_s$ runs) is extremely important as it would fix the values of various parameters in the Standard Model, providing bounds on new physics.

Computing the QCD running coupling is a major challenge for the lattice community. In this respect, several different methods have so far been investigated [1–4]. Such methods, excepting studies of the static potential, share the disadvantage that for each choice of the simulation parameters only one value of the running coupling can be determined. Thus, in general, several different data sets are needed in order to observe the scale dependence of the coupling. On the other hand, when extracting the coupling from the static potential, one has to fit the data to some phenomenological expression, so that the coupling is determined in a rather indirect way.

Here we investigate the feasibility of a more recent proposal [5]. It differs from the previous ones in that it adopts a definition of the coupling directly from the fundamental fields appearing in the continuum Lagrangian. Such a method allows the determination of the coupling at different scales from a single simulation, and also has the advantage that the extension to full QCD presents, in principle, no extra problems.

## 2. THE METHOD

Consider the standard lattice definition of the gluon fields in terms of the link variables [6],

$$A_\mu(x) = \left. \frac{U_\mu(x) - U_\mu^\dagger(x)}{2iag_0} \right|_{\text{traceless}} \quad (1)$$

where $a$ is the lattice spacing and $g_0$ the bare coupling. One can define the n-point gluon correlation functions in momentum space as

$$G^{(n)}_{\mu_1\ldots\mu_n}(p_1,\ldots,p_n) = \langle A_{\mu_1}(p_1)\ldots A_{\mu_n}(p_n)\rangle, \quad (2)$$

imposing momentum conservation. In particular,

$$G^{(2)}_{\mu\nu}(p) \equiv T_{\mu\nu}(p)\, G(p^2) \quad (3)$$

is the standard gluon propagator. The amputated three-gluon vertex is defined as

$$\Gamma^{(3)}_{\alpha\beta\gamma}(p_1,p_2,p_3) \equiv \frac{G^{(3)}_{\alpha\beta\gamma}(p_1,p_2,p_3)}{G^{(2)}(p_1)G^{(2)}(p_2)G^{(2)}(p_3)} \quad (4)$$

The vertex renormalisation constant, $Z_V(qa)$, can be defined at any fixed momentum scale $q$ by choosing appropriate kinematics to project out the part of $\Gamma$ proportional to the tree-level vertex [5]. We work in the Landau gauge, and noting the general form of $\Gamma$ in the continuum (in any covariant gauge [7]), we evaluate $\Gamma^{(3)}_{\alpha\beta\gamma}$ at the kinematical points defined by

$$\alpha = \gamma \neq \beta, \ \ p_1 = -p_3 = p_\beta, \ p_2 = 0. \quad (5)$$

---
[*]work done in collaboration with D.S. Henty.

Here, $p_\beta = q\ \hat{e}_\beta$, where $\hat{e}_\beta$ is the unit vector in the $\beta$ direction. Then, in the limit $a \to 0$, $\Gamma$ can be written as

$$\Gamma^{(3)}_{\alpha\beta\alpha}(-p_\beta, 0, p_\beta) = 2\ Z_V^{-1}(qa)\ g_0\ q. \quad (6)$$

Analogously, the wavefunction renormalisation constant $Z_A(qa)$ is defined from the gluon propagator by

$$G(q^2) = Z_A(qa)\ \frac{1}{q^2}. \quad (7)$$

The renormalised coupling $g(q)$ (where $\alpha_s(q) = g^2(q)/4\pi$) is finally obtained via

$$g(q) = Z_A^{3/2}(qa)\ Z_V^{-1}(qa)\ g_0. \quad (8)$$

From our definition, one can check explicitly that $g(q)$ is independent of $g_0$ (as it should be).

## 3. COMPUTATIONAL PROCEDURE

$SU(3)$ gauge configurations were produced using a hybrid-overrelaxed algorithm, every sixth sweep being a pseudo-heat-bath. We generated three data sets, each comprising 150 configurations, on a $16^4$ lattice. The values of $\beta = 6/g_0^2$ were chosen as 6.0, 6.2 and 6.4. Successive configurations were separated by 150 sweeps, 1000 sweeps being allowed for thermalisation.

A crucial step in the method is the accurate implementation of the lattice Landau gauge condition

$$\Delta(x) = \sum_\mu A_\mu(x + \hat{e}_\mu) - A_\mu(x) = 0, \ \ \forall x. \quad (9)$$

This is achieved by using a Fourier-accelerated algorithm [8]. To monitor the gauge-fixing accuracy we compute the quantity

$$\theta = \frac{1}{VN_C} \sum_x \mathrm{Tr}\ \Delta^\dagger(x)\Delta(x) \quad (10)$$

as the algorithm progresses, terminating when $\theta < 10^{-11}$ (this is close to 32-bit machine precision). Since our calculations involve low momentum modes of gluon correlation functions, the above test is not in itself sufficient since $\Delta(x)$ is a local quantity. For this reason, we also compute

$$A_0(t) = \sum_{\vec{x}} A_0(\vec{x}, t). \quad (11)$$

In a periodic box, the Landau gauge condition implies that $A_0(t)$ is independent of $t$ [6]. For our configurations, $A_0(t)$ is constant to better than one part in $10^5$.

For the purpose of our calculation, the only quantity which we need to store is the Fourier transform of $A_\mu(x)$ for a selected range of lattice momenta. All n-point gluon correlation functions can then be assembled at the analysis stage using Eq. 2, where momentum conservation is explicitly imposed. The calculation of the renormalised running coupling then proceeds as follows. First, we evaluate the gluon propagator and compute $Z_A$ from Eq. (7). Next, we compute the complete three-point function $G^{(3)}$ of the gluon field. We obtain $\Gamma$, the amputated vertex function, by dividing $G^{(3)}$ by the propagators corresponding to the momenta on the external legs. This yields $Z_V$ from Eq. (6). Finally, $g(q)$ is obtained from Eq. (8). We fully exploit the symmetries of Eq. (6) to increase statistics.

## 4. RESULTS

We plot the running coupling versus momentum in Fig. 1. We express $q$ in physical units using values for the inverse lattice spacing taken from string tension measurements [2,9].

We obtain a very clear signal despite the complicated numerical procedure. The important physical question is whether there exists a range of momenta for which the coupling runs according to the 2-loop perturbative expression

$$g^2(q) = \left[b_0 \ln(q^2/\Lambda^2) + \frac{b_1}{b_0} \ln\ln(q^2/\Lambda^2)\right]^{-1}. \quad (12)$$

Here, $b_0 = 11/16\pi^2$, $b_1 = 102/(16\pi^2)^2$ and $\Lambda$ is the QCD scale parameter for the pure gauge theory. To answer this question, and to obtain an estimate for $\Lambda$, we compute $\Lambda$ as a function of the measured values of $g(q)$ according to the formula

$$\Lambda = q\ \exp\left(-\frac{1}{2b_0 g^2(q)}\right) \left[b_0 g^2(q)\right]^{-\frac{b_1}{2b_0^2}}. \quad (13)$$

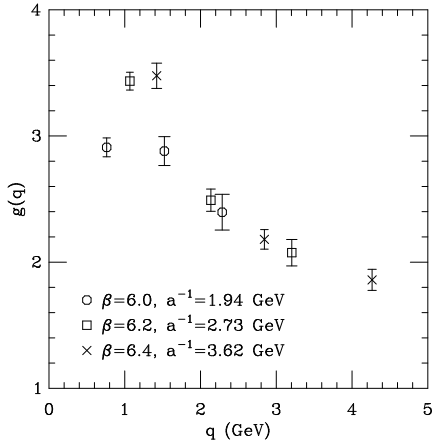

Figure 1. Running coupling vs. momentum.

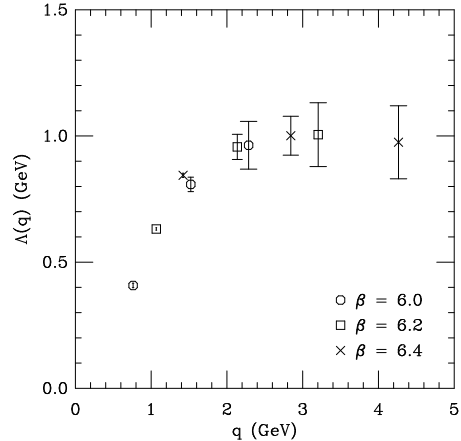

Figure 2. Lambda parameter vs. momentum.

If the coupling is running according to Eq. (12), then we expect $\Lambda$ defined from the above equation to be constant.

We plot $\Lambda$ versus $q$ in Fig. 2, and see that for $q > 2$ GeV the data are consistent with a constant value for the QCD scale parameter.

Since we have made measurements on different physical volumes and with different lattice momenta $qa$, we are able to check that, within errors, our results are independent of both the infra-red and ultra-violet cutoffs. We only show results obtained with at most three units of lattice momentum. As expected, when using higher momenta ($qa > 3\pi/8$), finite lattice spacing effects become clearly visible.

## 5. CONCLUSIONS

We have shown that a non-perturbative determination of the QCD running coupling can be obtained from first principles by a lattice study of the triple gluon vertex. Eventually, our values for the coupling $g(q)$ (or equivalently our value for $\Lambda$) will be converted to predictions for the corresponding quantities in the $\overline{\text{MS}}$ scheme via a matching calculation. Also, we plan to further analyse the role of lattice artifacts.

## 6. ACKNOWLEDGEMENTS

This work was carried out on the 16K Connection Machine 200 at the University of Edinburgh. The authors acknowledge the support of PPARC through grant GR/J 21347 (CP) and a Personal Fellowship (DH).

## REFERENCES

1. S.P. Booth et al., Phys. Lett. B294 (1992) 385.
2. G.S. Bali and K. Schilling, Phys. Rev. D47 (1993) 661.
3. A.X. El-Khadra et al., Phys. Rev. Lett. 69 (1992) 729.
4. M. Lüscher et al., Nucl. Phys. B413 (1994) 481.
5. C. Parrinello, Phys. Rev. D50 (1994) 4247.
6. J.E. Mandula, M. Ogilvie, Phys. Lett. B185 (1987) 127.
7. J.S. Ball, T.W. Chiu, Phys. Rev. D22 (1980) 2550.
8. C.T.H. Davies et al., Phys. Rev. D37 (1988) 1581.
9. C.R. Allton et.al., Phys. Lett. B284 (1992) 377.